\documentclass[11pt, a4paper]{article}
\usepackage{graphicx}
\usepackage{grffile}
\usepackage{longtable}
\usepackage{wrapfig}
\usepackage{rotating}
\usepackage[normalem]{ulem}
\usepackage{amsmath}
\usepackage{textcomp}
\usepackage{amssymb}
\usepackage{capt-of}
\usepackage{hyperref}
\usepackage{myPaperStyle}
\author[a,*]{D. D'Ambrosio}
\author[a,b]{J. Schoukens}
\author[a]{T. De Troyer}
\author[c]{M. Zivanovic}
\author[a]{M.C. Runacres}
\affil[a]{\footnotesize{Vrije Universiteit Brussel (VUB), Department of Engineering Technology (INDI), Pleinlaan 2, 1050 Brussels, Belgium.}}
\affil[b]{\footnotesize{Eindhoven University of Technology, Department of Electrical Engineering, Building Flux, P.O. Box 513, 5600 MB Eindhoven, The Netherlands.}}
\affil[c]{\footnotesize{Universidad Pública de Navarra, Electrical Engineering and Communication Department, Campus Arrosadia, s/n 31006 Pamplona, Spain.}}
\affil[*]{\footnotesize{\textit{Corresponding author. E-mail address: daniele.dambrosio@vub.be}}}
\date{}
\title{Phase-randomised Fourier transform model for the generation of synthetic wind speeds}
\hypersetup{
 pdfauthor={Daniele D'Ambrosio},
 pdftitle={Phase-randomised Fourier transform model for the generation of synthetic wind speeds},
 pdfkeywords={Synthetic wind speed, surrogate wind data, wind speed modelling, wind
  speed scenario, random-phase multisine},
 pdfsubject={},
 pdfcreator={Emacs 26.1 (Org mode 9.3.6)}, 
 pdflang={English}}
\begin{document}

\maketitle
\begin{abstract}
The increasing sophistication of wind turbine design and control generates a need for high-quality data. Therefore, the relatively limited set of measured wind data may be extended with computer-generated surrogate data, e.g.~to make reliable statistical studies of energy production and mechanical loads. This paper presents a data-driven, statistical model for the generation of realistic surrogate time series that is based on the phase-randomised Fourier transform. The proposed model simulates an ergodic, pseudo-random process that makes use of an iterative rank-reordering procedure to yield synthetic time series that possess the power spectral density of the target data and concurrently converges to the probability distribution of the target data with an arbitrary, user-defined precision. A comparison with two established data-driven modelling techniques for generating surrogate wind speeds is presented. The proposed model is tested under the same input conditions given in the test cases of the selected models, and its performance is investigated in terms of the agreement with the target statistical descriptors. Simulation results show that the proposed model can reproduce with high fidelity the statistical descriptors of the input datasets and is able to capture the nonstationary diurnal and seasonal variations of the wind speed.     
\end{abstract}
\keywords{Synthetic wind speed, surrogate wind data, wind speed modelling, wind
  speed scenario, random-phase multisine}
\vspace{1cm}

\section{Introduction}
\label{sec:org3f78211}
In recent years, the penetration of wind power in the electricity systems has
increased considerably \cite{lee2020global}. As a result, a growing need to
efficiently integrate the increasing share of wind energy into the grid has
emerged \cite{draxl_wind_2015}. The availability of high-quality wind-speed data
has become crucial to advance the integration process while keeping the cost of
wind energy low \cite{van_kuik_long-term_2016}. However, due to the cost and
duration of wind measurement campaigns it has become increasingly advantageous
to rely on surrogate wind data for the development of several strategic
applications. In particular, the latest advancements in power system modelling
with an increased share of wind energy
\cite{carapellucci_effect_2013,suomalainen_synthetic_2012}, in the design of
larger and lighter rotors as well as in control and condition monitoring
strategies have created a need for realistic surrogate time series of wind
speeds. The recent increase in the use of sonic anemometers as well as
reanalysis data such as MERRA-2
\cite{TheModernEraRetrospectiveAnalysisforResearchandApplicationsVersion2MERRA2}
has brought the advent of high-quality datasets that can be used to develop and
tune wind speed models for the generation of realistic wind-speed time series. 

The generation of surrogate times series is referred to in the meteorology and
wind energy communities as wind speed modelling, that is related to, but not
synonymous with, forecasting. The goal of wind speed modelling is not to predict
the future as in forecasting, but to computer-generate surrogate data that share
as many relevant features as possible with physical data. Wind time series are
characterised by probability density functions (PDF), expressing the relative
frequency of occurrence of wind speeds, and by power spectral densities (PSD)
or, equivalently, autocorrelation functions (ACF), expressing the temporal
coherence of the data. The PDF of wind speeds is typically non-Gaussian, unless
very short timespans are considered. The positively skewed Weibull distribution
is the most commonly used distribution for wind data \cite{CARTA2009933}. The
PSD characterises the wind time series in terms of the dominant frequencies and
the related temporal patterns that drive the wind-speed process
\cite{harris_macrometeorological_2008}. Moreover, wind time series are
inherently nonstationary as their PDF and PSD vary over time as a result of
deterministic meteorological factors changing over diurnal and seasonal time
scales \cite{barthelmie_observations_1996,dai_diurnal_1999,he_diurnal_2013}.
Therefore, capturing all the above-mentioned features poses a major challenge
for any wind speed model in the generation of realistic surrogate wind speed
time series. 

There exist different families of models to generate surrogate time series, more
or less common depending on the branch of physics or engineering where they are
applied. The first distinction to make is between physical modelling on the one
hand and data-driven (statistical) modelling on the other. Physical modelling generally involves the solution of physical conservation equations, as is done in numerical weather prediction tools such as WRF \cite{skamarock_description_2008}. Physical models are highly effective, but have a high computational cost and may produce more information than is required. Data-driven models on the other hand, do not to involve the solution of physical conservation laws, although they may use physical constraints on the model output, and are comparatively cheap. All data-driven models start from a set of training data and learn from those data how to produce surrogate time series with the same characteristics as the training data.

In the data-driven or statistical wind modelling class, established methods for
generating surrogate wind speeds broadly fall in one of two categories:
Markov-chain (MC) based models, and autoregressive integrated moving average
(ARIMA) models. Markov chain models do not require the estimation of a
continuous wind speed distribution and as such can conform to a discrete
measured distribution. The fact that they are capable of reproducing the
probability density function of the real data is indeed one of their main
advantages \cite{4840265}. Their main drawback is that their memory is limited
by the order of the model. As a consequence, daily trends can only be modelled
with time steps of a few hours. Equivalently, the 10 minute time steps
prescribed by the IEC-61400-12 standard \cite{IEC61400-12-1:2017} only allow
very short-term trends to be modelled. There is a practical limit to expanding the time memory of Markov chain models by adding extra orders, as the models quickly become prohibitively expensive with increasing model order. Nesting Markov chain models can improve the performance over standard Markov chain models but even for those models the autocorrelation quality quickly deteriorates as the time lag becomes larger \cite{TAGLIAFERRI2016118}.

 ARIMA models consist in a modified form of the autoregressive-moving average (ARMA) process, that are designed to model time series with a homogeneous nonstationary behavior \cite{box2015time}. In its standard modelling procedure, it is not well suited to model highly nonstationary and non-Gaussian random processes as the wind-speed variation and thus requires modifications to produce satisfactory simulation results \cite{Yunus2016}. A proper power transformation of the data can be introduced to partly overcome the limitation to Gaussian processes; with the further introduction of limitation and seasonal partition of the data, it has been shown to adequately model the monthly variation of wind power generation \cite{5340622}. Furthermore, ARIMA models struggle to capture reliably the diurnal and seasonal variations of wind
speeds. This can be partly mitigated by using nested ARIMA methods but even then the distributions of the wind data are far from perfect \cite{Sim2019}. Even with improvements such as nesting, neither the ARIMA nor the Markov chain models are fully satisfactory for wind modelling, as manifested in their inability to conform to both the PDF and the PSD (or the ACF) of a measured dataset. 

This paper presents a phase-randomised Fourier transform (PRFT) model for the generation of realistic surrogate time series of wind speeds. This class of data-driven models originated in the 1990s in the fields of nonlinear physics \cite{THEILER199277,schreiber_improved_1996} and system identification \cite{schoukens_design_1998}, and is able to produce surrogate data that do conform to both a given generally non-Gaussian PDF and a PSD. Futhermore, thanks to their iterative, rank-reordering process, PRFT models are ideal candidates to be tested for their potential to capture nonstationary features of the wind speed such as its diurnal and seasonal variations. To our knowledge, this class of models is not well established in the wind energy community and no attempt has been reported in the literature to apply them to the generation of surrogate wind data. To show its performance, the proposed PRFT model is tested and compared with two state-of-the-art statistical models for the generation of surrogate wind data that represent recent developments in the Markov chain and the ARIMA-based modelling category, respectively. The proposed model is applied to the same datasets used for the test cases of the selected models, and the simulation results are compared in terms of the accuracy in reproducing the PDF, the PSD, and the nonstationary features of the input data. The level of user interaction required by the selected models is also discussed.                
\section{Phase-randomised Fourier transform model}
\label{sec:org9e834a9}
\begin{figure*}
\centering
\includegraphics[width=\textwidth]{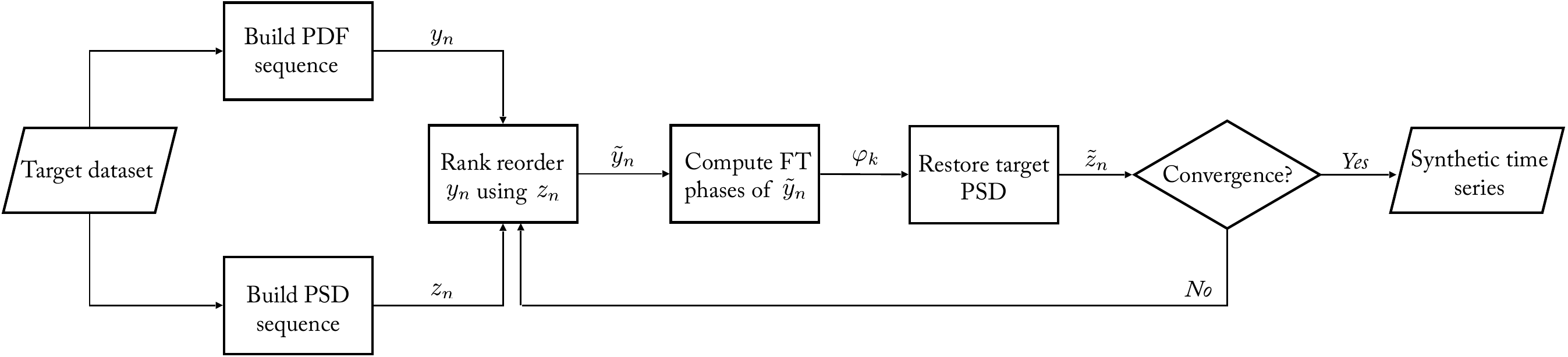}
\caption{\label{fig:org16dbe28}High-level flowchart of the proposed PRFT model.}
\end{figure*}

The proposed PRFT model is presented in this section and its algorithm is illustrated in the high-level flow chart of Fig.~\ref{fig:org16dbe28}. In the initialization phase, two initial sequences are generated that are consistent with the PSD and the PDF of the target wind-speed time series, respectively. Next, an iterative process first reorders the sequence conforming to the target PDF to match the rank order of the sequence possessing the target PSD. Then, a new sequence is generated from the Fourier amplitudes of the target data and the spectral phases of the last reordered sequence. This new sequence replaces the initial sequence with the target PSD in the next reordering step, and the iterations continue until the last generated sequence has converged to the PDF of the target data.  

\subsection{Initialization phase}
\label{sec:orgd9312b2}
The initial sequence possessing the target PSD is generated as a random-phase multisine signal from the frequency domain. This is achieved by performing an inverse Fourier transform on the Fourier amplitudes of the target data multiplied by a set of phases \(\varphi_{k}^{\text{rnd}}\) randomly sampled from the uniform distribution \(\mathcal{U} [0, 2\pi)\). Let \(x(t_n)\) be the measured wind-speed time series from which one wants to generate synthetic data that consist of \(N\) observations taken at the equally-spaced times \(t_{n} = n\Delta t\), \(n = 0, \dots, N-1\), where \(\Delta t\) is the sampling interval or averaging time of the target time series. The absolute values of its Fourier amplitudes are calculated via the discrete Fourier transform
\begin{equation}
  |X(f_k)| = |\mathcal{F}\{ x(t_n)\}| = \left| \sum_{n=0}^{N-1} x(t_{n})e^{-2\pi j f_{k}n\Delta t}\right|,
\end{equation}
that are estimated at the \(N\) discrete frequencies \(f_{k} = k\Delta f\), \(k = 0, \dots, N-1\), dictated by the sampling frequency of the target time series, \(f_{s} = 1/\Delta t\), according to the definition of the frequency resolution, \(\Delta f = f_{s}/N\). Then, the mean of the target time series is removed by setting to zero the Fourier amplitude corresponding to the zero-frequency, \(X_{0}= 0\), and only the values estimated at the frequencies \(f_{0}, \dots, f_{N/2-1}\) are retained by performing zero padding so as to get \(|X(f_k)| = 0\) for \(k = N/2, \dots, N-1\). Finally, the time-domain initial sequence possessing the same PSD of the target data is generated by inverse Fourier transforming the resulting \(|X(f_k)|\) multiplied by the random phases \(\varphi_{k}^{\text{rnd}}\),
\begin{equation}
 z_{n} = 2\Re\{ \mathcal{F}^{-1}\{ |X(f_k)|e^{j\varphi_{k}^{\text{rnd}}}\} \},
\end{equation}
where \(\Re\{\cdot\}\) indicates the real part.

The initial sequence consistent with the PDF of the target wind-speed dataset is obtained by performing an inverse CDF transform on an equally spaced sequence of \(N\) samples taken on the interval \((0,1)\), and denoted by \(u_{n}\). This sequence represents the cumulative distribution function of a uniformly distributed variable and is defined as
\begin{equation}
u_{n} = \frac{1+2n}{2N},\:\:\:\:n = 0, ..., N-1.
\end{equation}
This technique allows to obtain a sequence of \(N\) samples distributed with the desired probability density function by performing
\begin{equation}
y_{n} = F^{-1}(u_{n}),
\label{eq:inverseCDF}
\end{equation}
where \(F^{-1}\) is the inverse target CDF. Such a technique is not restricted to cumulative distribution functions that possess an analytical inverse. In cases where \(F^{-1}\) can not be analytically determined, a nonlinear curve fitting can be performed to obtain a numerical inverse of the target CDF. Moreover, Eq.~\eqref{eq:inverseCDF} allows to directly implement the empirical cumulative distribution function of the target wind-speed dataset if one wishes to retrieve a better match with the observed PDF without performing a numerical fit to the probability distribution of the wind-speed data.

To complete the initialization phase, the variance and the mean of the sequence consistent with the target PDF, \(y_{n}\), are imposed to \(z_{n}\), and the absolute values of the Fourier amplitudes of this sequence are stored, \(Z_{k} = |\mathcal{F}\{z_{n}\}|\).
\subsection{Iterative process}
\label{sec:org0aebf84}
At this stage, an iterative process is implemented to get one sequence that is consistent with both the target PSD and PDF of the wind-speed data. This process consists of three steps (Fig.~\ref{fig:org16dbe28}). Firstly, a rank-reorder step is performed through a non-linear transformation that reorders \(y_{n}\) in a new sequence \(\tilde{y}_{n}\) so that the smallest value of \(y_{n}\) is given the same position in \(\tilde{y}_{n}\) that the smallest value of \(z_{n}\) has in its own sequence, and so forth for all the \(N\) values. In the second step, the phases of the Fourier spectrum of this new sequence \(\tilde{y}_{n}\) are computed by means of the Fourier transform of the signal, which yields
\begin{equation}
\varphi_{k} = \tan^{-1}\left(\frac{\Im\{\mathcal{F}\{\tilde{y}_{n}\}\}}{\Re\{\mathcal{F}\{\tilde{y}_{n}\}\}}\right),
\label{eq:phases}
\end{equation}
where \(\Im\{\cdot\}\) and \(\Re\{\cdot\}\) indicate the imaginary part and the real part, respectively. In the last step, the final sequence \(\tilde{z}_{n}\) is generated by inverse Fourier transforming the stored Fourier amplitudes \(Z_{k}\) multiplied by the last computed phases \(\varphi_{k}\), and then taking the real part of the inverse Fourier transform; that is   
\begin{equation}
\tilde{z}_{n} = \Re\{\mathcal{F}^{-1}\{Z_{k} e^{ j \varphi_{k}}\}\}.
\label{eq:ifftlast}
\end{equation}
This last generated sequence replaces \(z_{n}\) in the successive rank-reordering step of \(y_{n}\) performed in the next iteration, and this three-step process is iterated until the probability density function of \(\tilde{z}_{n}\) converges to the PDF of the target wind-speed dataset.

This iterative process is in essence equivalent to the iterative scheme of the iterative amplitude adjusted Fourier transform (IAAFT) algorithm proposed by Schreiber and Schmitz \cite{schreiber_improved_1996}. Its main difference resides in the order in which the steps are carried out. In the proposed scheme, the rank-reorder is performed first and the imposition of the target PSD is carried out as the last step; this order is reversed in the IAAFT algorithm. Therefore, the final signal delivered by the proposed algorithm is the one after the restoration of the target PSD, whereas in the IAAFT approach it is the one produced by the last rank-reorder.

\subsection{Model properties}
\label{sec:org205441b}
By its design, the proposed PRFT model simulates an ergodic, pseudo-random process. This emphasises two useful features of the model. First, each realisation of the simulated process results in a synthetic time series that always conforms to the same statistical descriptors which are, by construction of the model, the PDF and the PSD of the input wind data. Second, the initial random phases \(\varphi_{k}^{\text{rnd}}\) give rise to a stochastic reordering of the synthetic wind speeds without altering their value in a random fashion, and that effectively delivers the same synthetic wind speeds with a different time evolution for each realisation of the model.       

As for the extreme wind speeds, repeating the simulation with the same input data and with the same sampling defined by Eq.~\eqref{eq:inverseCDF} results in a different synthetic signal characterized by the same extreme values. These extremes, however, appear at different time instants in the synthetic signal as a result of the different random seed drawn for the inital phases \(\varphi_{k}^{\text{rnd}}\). Therefore, if one wishes to extend the simulated extreme winds, the number of the simulated samples \(N\) is to be increased so as to increase the sampling in the tail region of the target CDF.

Periodic features of the target data such as the diurnal and seasonal variations create peaks in the PSD. The number of components in Eq.~\eqref{eq:ifftlast} that is needed to model these periodic variations is minimized if an integer number of years (the seaonal variations) and days (the diurnal variations) are included in the synthesized signal. This leads to a sparse representation (only a few parameters are needed) of the periodic phenomena in the synthetic data that better mimics the observed seasonal/diurnal behavior in the target data.
\section{Comparison with MC and ARIMA models}
\label{sec:orgface37b}
The performance of the proposed PRFT model is compared with two state-of-the-art
modelling techniques for the simulation of nonstationary wind speeds, namely the
non-homogeneous Markov chain model (NHMC) put forward by Xie et
al.~\cite{Xie2017}, and the ARIMA-based frequency-decomposed methodology
presented by Yunus et al.~\cite{Yunus2016}. These models are selected as a benchmark since both aim to reproduce the probability distribution and the time correlation of the observed data, while attempting to capture seasonal and diurnal variations that are characteristic of the wind-speed stochastic process. Such nonstationary features are taken into account by adopting different modelling strategies.

In the NHMC model, the time homogeneity assumption is relaxed allowing the Markov chain transition probability matrix to become a function of time. Then, the time-varying transition matrix, that represents the wind speed variation at different times, is generated by means of a seasonal partition technique and a sequence period extraction procedure performed on the wind data. This allows to yield state transition probabilities that are time related and fine adjusted to simulate the diurnal and seasonal variation of the wind speed.

With regard to the ARIMA-based model, a frequency decomposition of the wind-speed data is introduced in the standard ARIMA modelling procedure \cite{box2015time} to better capture the periodic and nonstationary characteristics of the wind-speed fluctuations. This decomposition allows to split the wind-speed data into a high-frequency (HF), stationary component, and a low-frequency (LF), nonstationary component that accounts for the seasonal and diurnal cyclical variation of the wind speed. Both components are in turn modelled separately by performing a standard ARIMA procedure, and then combined to get the synthetic wind-speed time series. In addition, shifting and limitation of the wind-speed data are introduced before modelling and during simulation with respect to the standard ARIMA modellling procedure.     

The comparison carried out in this section aims at showing the performance of the proposed PRFT model in reproducing the PDF and the PSD, or equivalently, the autocorrelation function of the observed wind data when it is applied to the same datasets of the test cases presented in Xie et al.~and Yunus et al.~for the NHMC model and the ARIMA-based model, respectively. Particular emphasis is put on how accurately the proposed model can capture the nonstationarity of the wind-speed data pertaining to the seasonal and diurnal variation of the wind speed. In addition, the level of user interaction required by the selected models is discussed.                       
\subsection{NHMC model comparison}
\label{sec:org7778829}
The proposed PRFT method is applied to the same dataset used in the case study
of Xie et al.~\cite{Xie2017}, recorded at Crosby USA, that is the hourly
averaged wind speeds extracted from the database available on the internet at
the North Dakota Agricultural Weather Network (NDAWN) website \cite{ndawn} for the 10-year period ranging from January 2003 to December 2012. A synthetic wind-speed time series of the same length is generated, and the agreement of the proposed model with the target data is investigated in terms of its probability distribution, its power spectral density, and its autocorrelation function. To allow for a direct comparison, the same statistical descriptors produced by the NHMC model and shown in the Xie et al.~case study are digitised and presented along with the results given by the PRFT model. In addition, the degree of nonstationarity reproduced in the synthetic data is shown and compared in terms of the average seasonal variation (ASV) of the wind speed, and the amplitudes of the diurnal cycle harmonics of the wind speed detected in the PSD.

\subsubsection{Average Seasonal Variation}
\label{sec:org1020d90}
\begin{figure}[htbp]
\centering
\includegraphics[width=0.45\textwidth]{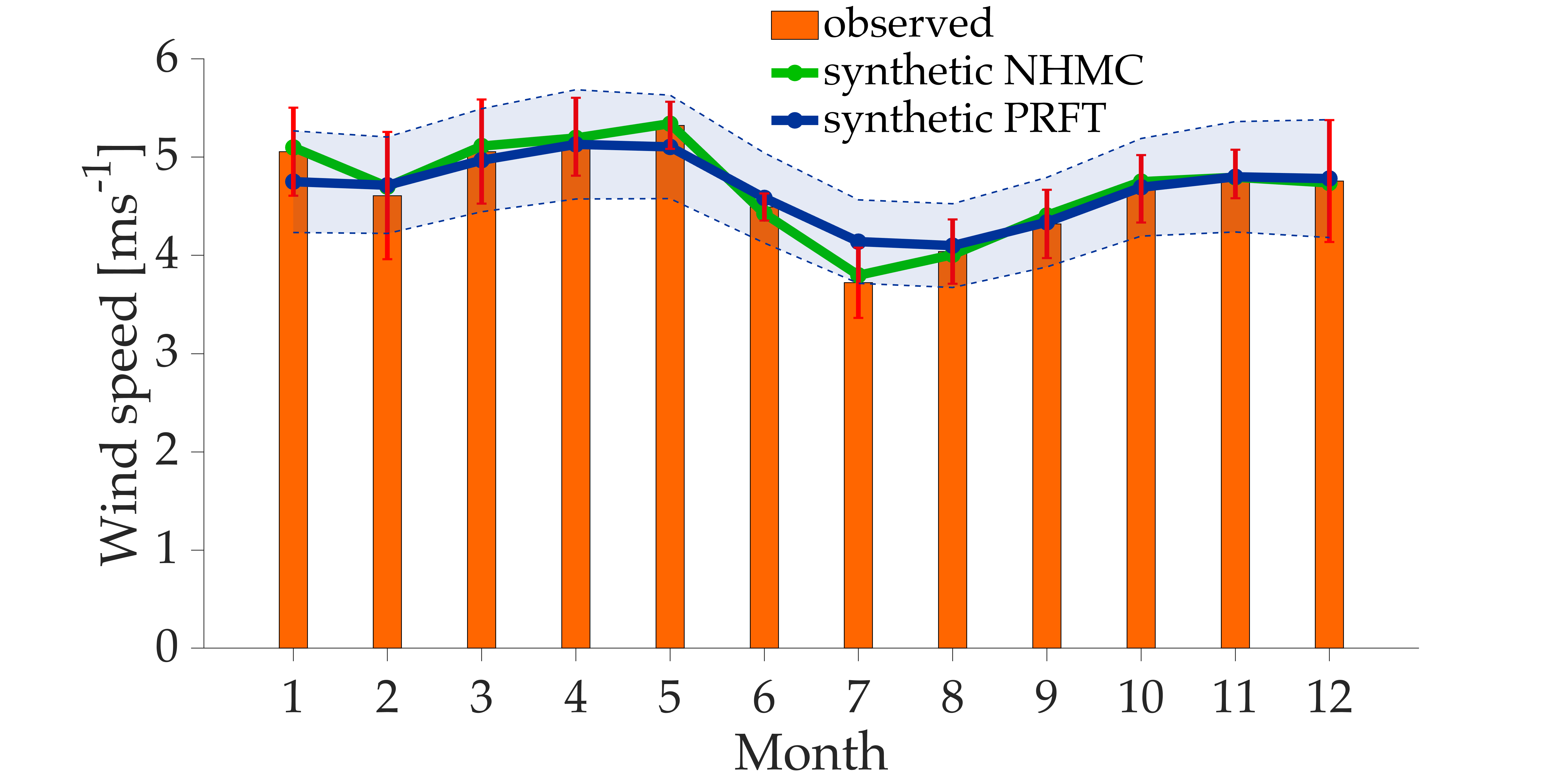}
\caption{\label{fig:org0d52db7}Average seasonal variation comparison shown as monthly-average wind speeds of 10-year data. ASV observed at Crosby as orange bars along with its inter-annual variability as red error bars; PRFT-generated ASV as blue, thick line with its confidence intervals as blue, shaded regions (average over 10000 simulations); NHMC-generated ASV as green, thick line.}
\end{figure}

Fig.~\ref{fig:org0d52db7} shows the ASV of the observed wind speeds at Crosby calculated as the monthly-average wind speed across the 10 years of data (orange bars), along with its inter-annual variability calculated as the associated standard deviation (red error bars). The same average variation is shown for the synthetic data simulated by the PRFT model (blue thick line) and by the NHMC model (green thick line). The ASV shown for the PRFT model is an average result obtained over 10000 realizations of the model; the confidence strips of the multiple realizations are calculated as the associated standard deviation (blue shaded regions), and indicate the inter-annual variability of the monthly-average synthetic data. This is done as the initial random seed for the phases \(\varphi_{k}^{\text{rnd}}\) determines a stochastic rearrangement of the synthetic wind speeds that ultimately leads to an ASV varying in a limited range with the different realizations of the model. 

The monthly fluctuation observed in Fig.~\ref{fig:org0d52db7} reveals a significant intra-annual or seasonal variation of the wind speed in the observed data at Crosby. This seasonal variation is well reproduced by the proposed PRFT model, whose ASV averaged across multiple simulations is consistent with the ASV of the observed wind data. From a visual comparison with the ASV produced by the NHMC model it can be noticed that overall the PRFT model shows a comparable performance with the Xie et al.~model in simulating the observed seasonal variation at Crosby. For some months, the PRFT model yields slightly larger deviations from the observed data compared to the NHMC model (January, May, and July), showing a poorer performance. However, note that the PRFT-simulated ASV is presented here as an average over multiple realisations to show its convergence to the observed data; when simulating, the ASV error can be computed and one may reject simulation results until a prescribed tolerance is satisfied. In contrast, Xie et al.~do not specify whether their simulation result represents one realisation of the NHMC model or is an average over multiple realisations. Nevertheless, a degree of randomness in the simulation results of the NHMC model is to be expected as Markov chain methods belong to the class of Monte Carlo methods.      
\subsubsection{Probability Distribution}
\label{sec:orgcef262f}
\begin{figure}[htbp]
\centering
\includegraphics[width=0.45\textwidth]{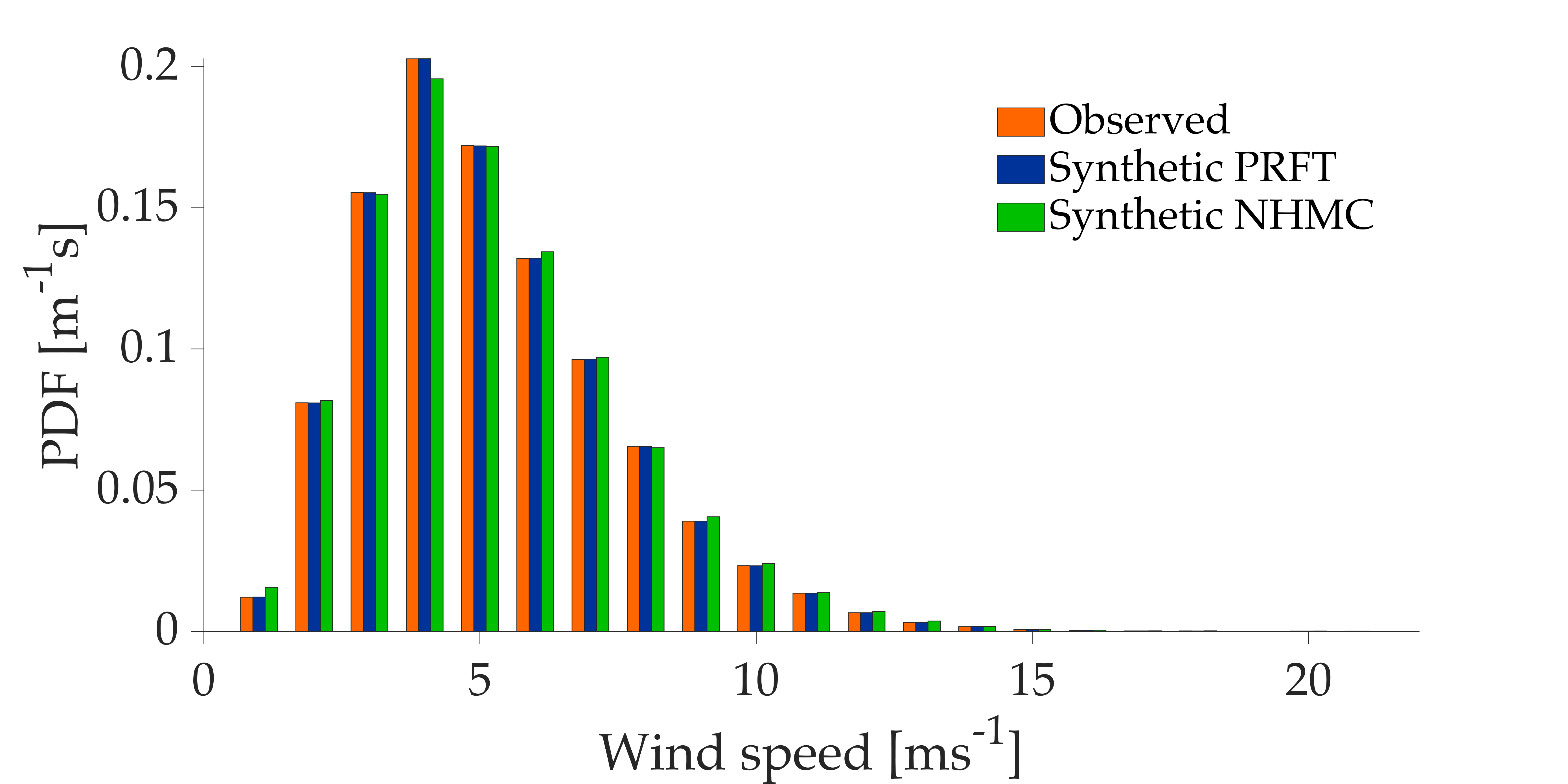}
\caption{\label{fig:org80411c2}PDF agreement of synthetic wind speeds generated by the PRFT model and by the NHMC model with the wind data observed at Crosby .}
\end{figure}

The PDF of the synthetic wind speeds generated by the proposed model is shown in Fig.~\ref{fig:org80411c2} along with the probability distributions of the observed wind speeds at Crosby and the synthetic speeds produced by the NHMC model. It can be noticed that the PDF yielded by the PRFT model is consistent with the observed wind-speed distribution, and its simulated wind speeds fit very accurately the target stochastic process. The goodness of the fit with the observed wind data is evaluated in terms of the \(R^{2}\) coefficient and the root-mean-square error of the cumulative distribution function, \emph{RMSE}\(_{\text{CDF}}\). In Table \ref{tab:org94502f1}, these statistics are summarised and compared with the values obtained from the application of the NHMC model and provided in Xie et al. The visual comparison and the reported metrics show that the PFRT model and the NHMC model attain a similar performance in reproducing the probability distribution of the wind speed measured at Crosby.

\begin{table}[htbp]
\caption{\label{tab:org94502f1}Goodness of fit of the compared models for the Crosby dataset.}
\centering
\begin{tabular}{|c|c|c|c|}
\hline
Site & Model & \(R^{2}\) & \emph{RMSE}\(_{\text{CDF}}\)\\
\hline
\hline
Crosby &  &  & \\
 & PRFT & 0.999999 & 0.0005\\
 & NHMC & 0.999963 & 0.0020\\
 &  &  & \\
\hline
\end{tabular}
\end{table}
It is important to note that the observed PDF is always reproduced with the same level of accuracy when multiple simulations are carried out with the proposed PRFT model. This means that the random phases drawn to generate the initial PSD sequence (random-phase multisine) do not affect the values of the generated wind speeds but only determine their reordering in the time series.       
\subsubsection{Autocorrelation Function}
\label{sec:orgfa3b1b5}
\begin{figure}[htbp]
\centering
\includegraphics[width=0.45\textwidth]{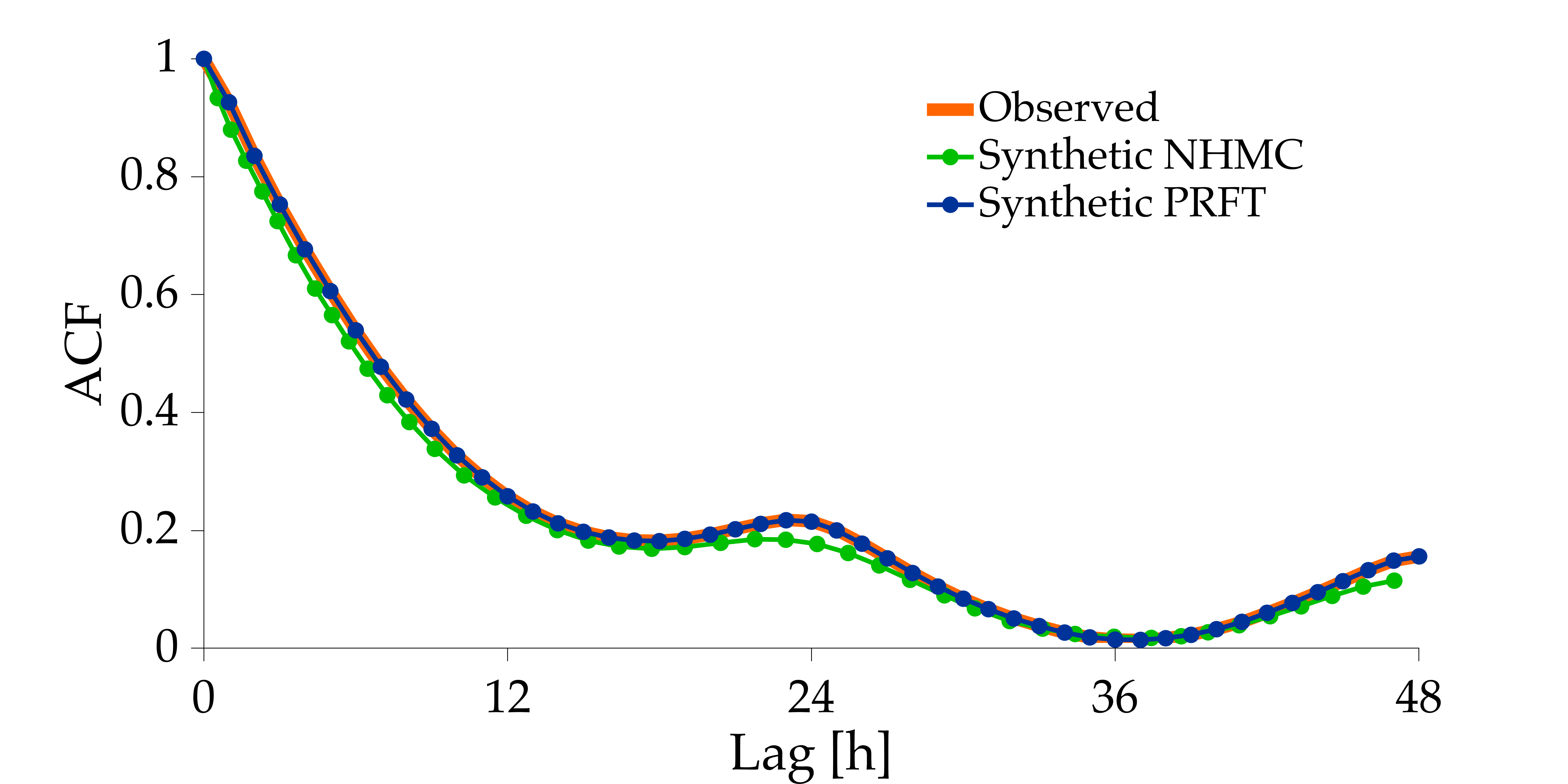}
\caption{\label{fig:orgb27b88d}ACF agreement of synthetic wind speeds obtained from the PRFT model and from the NHMC model with wind data measured at Crosby.}
\end{figure}

The analysis of the autocorrelation function of the wind speed modelled by the proposed PRFT model resulted in the agreement shown in Fig.~\ref{fig:orgb27b88d}. The synthetic ACF is presented for the first 48 lags, corresponding to 48 hours, along with the observed ACF at Crosby, and it is compared with the autocorrelation function produced by the NHMC model for the same number of lags. In addition, the root-mean-square error of the ACF, \emph{RMSE}\(_{\text{ACF}}\), is calculated for four selected time lags and compared in Table \ref{tab:org8bf34b4} with the same ACF error produced by the NHMC model and presented in the case study of Xie et al.   

\begin{table}[htbp]
\caption{\label{tab:org8bf34b4}ACF error produced by the compared models for different lags when synthetizing from the Crosby dataset.}
\centering
\begin{tabular}{|c|c|c|c|c|c|}
\hline
Model & 12-h Lag & 24-h Lag & 48-h Lag & 100-h Lag\\
\hline
\hline
 &  &  &  & \\
PRFT & 0.0001 & 0.0001 & 0.0001 & 0.0001\\
NHMC & 0.0247 & 0.0193 & 0.0160 & 0.0157\\
 &  &  &  & \\
\hline
\end{tabular}
\end{table}
This analysis shows that the proposed PRFT model generates synthetic wind speeds whose ACF is consistent with the autocorrelation function of the wind speed observed at Crosby. Fig.~\ref{fig:orgb27b88d} illustrates that the proposed model can reproduce accurately the diurnal correlation of the observed wind speed indicated by the periodic variation of the ACF with a 24-hour period. Also, a visual comparison reveals that the NHMC model yields a similar agreement with the observed ACF, being able to capture both the ACF decaying trend and the diurnal correlation structure of the wind speed at Crosby. From the quantitative comparison presented in Table \ref{tab:org8bf34b4}, it can be noticed that the PRFT model yields a RMSE in the ACF that is consistently lower than the ACF error produced by the NHMC model for all the analysed time lags.

When multiple simulations are performed with the PRFT model, the target ACF is always reproduced with the same level of accuracy as the initial random phases do not affect the autocorrelation structure of the simulated process. 
\subsubsection{Power Spectral Density}
\label{sec:orgf8daec1}
\begin{figure}[htbp]
\centering
\includegraphics[width=0.45\textwidth]{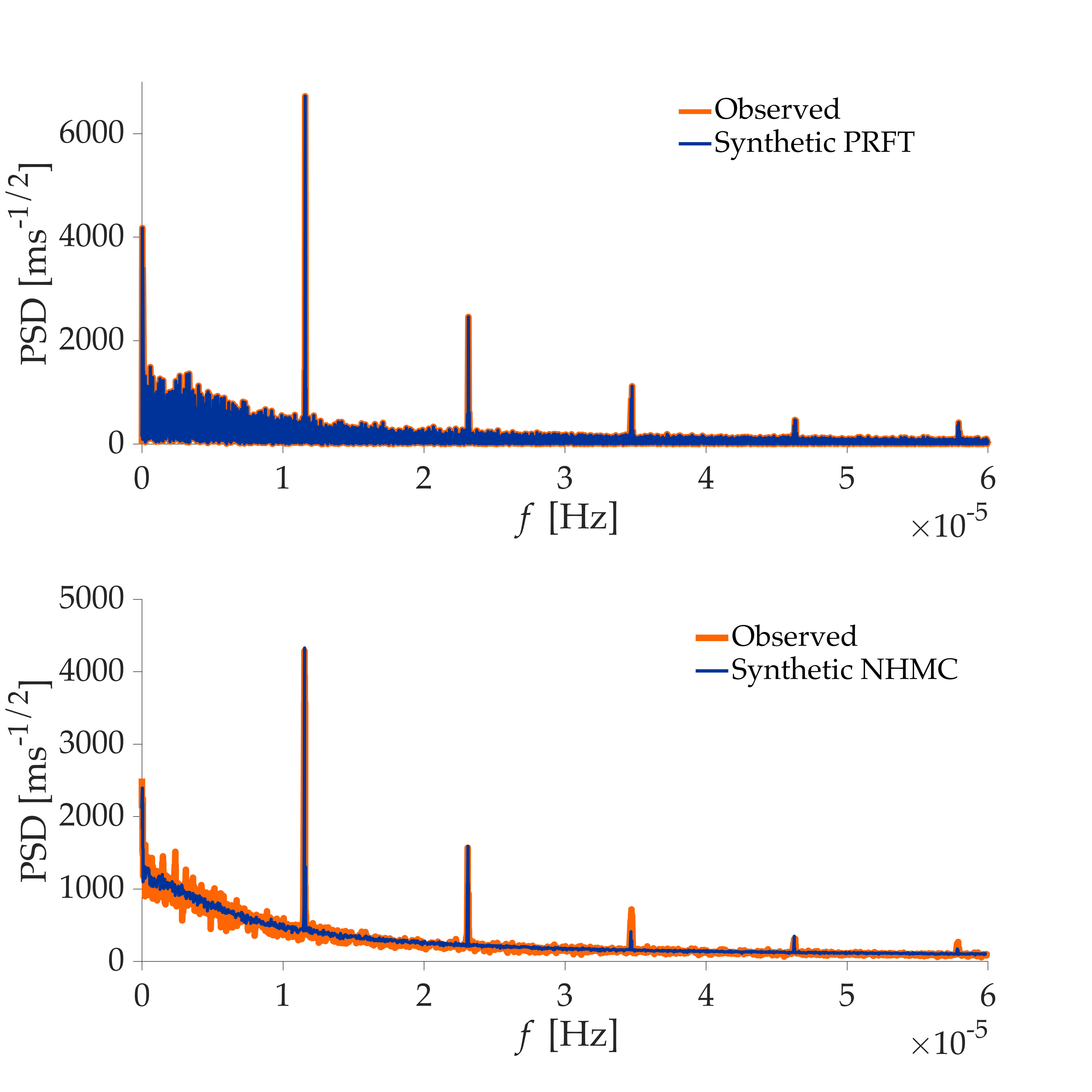}
\caption{\label{fig:orga89d162}Square root PSD agreement of synthetised wind speeds generated by the PRFT model (\emph{top}) and by the NHMC model (\emph{bottom}) with observed wind data at Crosby.}
\end{figure}

The performance of the proposed PRFT model in reproducing the spectral energy content of the observed wind speeds at Crosby is investigated through a spectral analysis of the simulated wind speeds. This analysis allows to reveal the presence of deterministic frequency components that give rise to nonstationarity in the wind-speed data, and to assess how accurately these components are simulated by the synthetic data. To this end, the PSD of the PRFT-simulated wind speeds is shown in the top plot of Fig.~\ref{fig:orga89d162}, along with the PSD of the observed wind speeds at Crosby. To allow for a direct comparison with simulation results given in Xie et al., the PSD produced by the NHMC model is digitised and presented in the bottom plot of the same figure. The simulation results of Xie et al.~are presented in a separate plot as the PSD of the observed data shown there does not coincide with the PSD calculated from the observed wind speeds and shown in the top plot of Fig.~\ref{fig:orga89d162}. This suggests that some technique was applied to reduce the variance of the estimated PSD of the observed data, and the same technique is likely to have been applied to the NHMC-simulated PSD. As the digitised PSD is given in ms\(^{-1/2}\), the PSD obtained from the PRFT model is shown in the same units by taking the square root of its values. For visualisation purposes, the zero-frequency or DC component of the PSD is not shown, and the lowest spectral peak is the annual frequency component occurring at \(f\) = 3.171 \(\times\) 10\(^{-8}\) Hz.  

The PSD analysis shows that the PRFT model can reproduce with high accuracy the entire frequency content of the wind speed at Crosby, including the strongest deterministic components detected in the observed data at the annual, diurnal and semidiurnal time scales, namely at \(f\) = 3.171 \(\times\) 10\(^{-8}\), 1.157 \(\times\) 10\(^{-5}\), and 2.314 \(\times\) 10\(^{-5}\) Hz, respectively. The goodness of the agreement with the observed PSD is calculated as the root-mean-square error in the PSD produced by the PRFT model, \emph{RMSE}\(_{\text{PSD}}\), that results in a value of 0.00021. The same level of accuracy in reproducing the observed PSD is attained for multiple realizations of the PRFT model, as the stochastic reordering due to the initial random phases \(\varphi_{k}^{\text{rnd}}\) always yields a synthetic wind-speed time series consistent with the target PSD. Note that the presence of a strong diurnal component in the frequency domain is associated with the 24-hour periodic variation observed for the ACF, as the power spectrum and the aucorrelation function constitute a Fourier-transform pair according to the Wiener-Khinchin theorem \cite{wiener1949time}.           

A visual inspection of the PSD agreement produced by the NHMC model (bottom plot of Fig.~\ref{fig:orga89d162}) reveals that the NHMC-simulated PSD fails in reproducing the spectral content of the Crosby data with the same level of accuracy yielded by the PRFT model. The NHMC model underperforms in estimating the amplitudes of the observed PSD at Crosby throughout the analysed frequency range, with the exception of the diurnal and semidiurnal harmonics whose amplitudes are correctly captured by the model. Larger deviations are observed for the amplitudes of the 3rd and the 5th harmonics of the diurnal cyle. Also, the NHMC model fails in adequately reproducing the frequency content between the annual and the diurnal peak, that represents the wind speed fluctuations associated with the passage of large, synoptic-scale pressure systems \cite{vanderhoven1957}. As Xie et al.~do not provide any metric for the deviations from the observed PSD, it is not possible to perform a quantitative comparison with the error in the PSD obtained from the PRFT model. 
\subsubsection{User Interaction}
\label{sec:org761fc56}
As described in Xie et al., the NHMC model construction requires user tuning during the preprocessing phase in order to enable the modelling of the seasonal and diurnal characteristics of the specific input wind data. In the seasonal effect partition, an optimal partition method is implemented to split the wind data in a number of segments that reflect the seasonal variability of the data. In this step, a user choice is required to determine the optimal number of segments as this is not specified by the optimal partition method. Then, after performing the sequence period extraction, the user has to decide on the most suitable wind variation period \(R\) according to the periodic characteristics of the wind data; this parameter will determine in turn the number of transition probability matrices used by the model to represent the wind speed variation at different times. Finally, any Markov chain model requires an initial choice by the user on the number of states into which the input wind data are discretised, that define the state transition probabilities of the transition matrix. Overall, the user interaction required during the preprocessing phase affects the performance of the NHMC model and necessitates some expertise by the user to fine tune the NHMC model.

In contrast, the PRFT model is fully automated and no tuning of the model is required to generate synthetic data from target data with any length and temporal resolution. It is only required that the target data contain an integer number of days and years so as to avoid introducing any leakage in the periodic components of the PSD.      
\subsection{ARIMA model comparison}
\label{sec:org253ff64}
The ARIMA-based frequency-decomposed model put forward by Yunus et al.~\cite{Yunus2016} is the second model selected for benchmarking the proposed methodology. To do that, the PRFT model is applied to generate synthetic wind data from the same dataset used in their test case, namely the 10-minute average wind speeds recorded by the meteorological mast located in the Näsudden peninsula in Gotland, Sweden, from the 1st of January to the 31st of December 2005 at a height of 100 m. This dataset is maintained and provided by the Department of Earth Sciences of Uppsala University. A synthetic wind-speed time series of the same length of the dataset is generated, and the comparison with the ARIMA-simulated wind data is carried out in terms of the same statistical descriptors presented in Section IV of Yunus et al., namely the probability distribution, the autocorrelation function, and the power spectrum. Their values are digitised and shown along with the results given by the application of the proposed PRFT model. 

\subsubsection{Probability Distribution}
\label{sec:org1c42b8b}
The performance of the two models is first compared in terms of their capability to reproduce the probability distribution of the observed wind speeds. A quantile-quantile (Q-Q) plot is employed to assess such a capability. This type of plot provides a graphical method to compare two PDFs: the closer the data points lay on the straight line \(y = x\), the better is the agreement of the compared probability distributions.

\begin{figure}[htbp]
\centering
\includegraphics[width=0.4\textwidth]{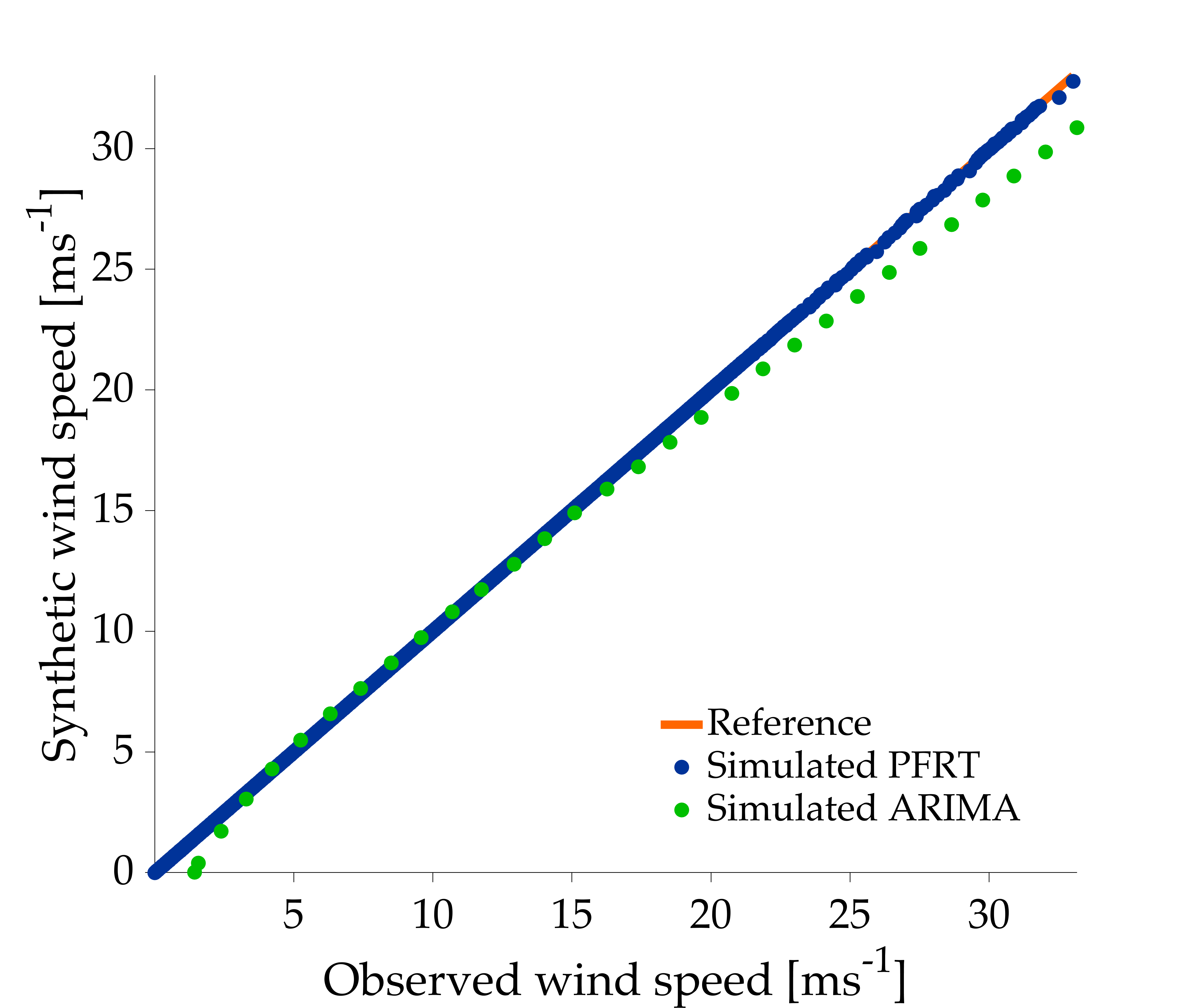}
\caption{\label{fig:orgfd0f992}Quantile-quantile plot of observed wind speeds at Näsudden against synthetic wind speeds generated by the PRFT model and the ARIMA-based model.}
\end{figure} 
Fig.~\ref{fig:orgfd0f992} shows the Q-Q plot of the synthetic wind speeds obtained from the application of the proposed PRFT model, along with the digitised Q-Q plot produced by the ARIMA-based modelling. It can be noticed that the wind data simulated with the PRFT model are in very good agreement with the observed wind speeds at Näsudden. In contrast, the Q-Q plot given by the ARIMA-simulated data reveals deviations from the reference data occurring at wind speeds around 15 ms\(^{-1}\) that become larger with increasing wind speed. Additionally, significant deviations of the ARIMA-simulated data from the observed data can be also observed in the very low wind-speed region between 0 and 3 ms\(^{-1}\). Yunus et al.~comment that the underperformance of the ARIMA-model in reproducing the largest wind speeds is due to the limited number of wind speeds higher than 20 ms\(^{-1}\) in the Näsudden dataset. They further comment that the extreme wind conditions can be properly modelled with a separate technique when required.

In contrast, the extreme wind speeds observed at Näsudden are well captured by the PRFT model, that shows a very limited deviation with respect to the observed data for the highest wind-speed occurrencies. Moreover, the kernel of the proposed model guarantees that the same extreme values are generated for each realisation of the PRFT model when it is applied to the same input data and with the same sampling defined by Eq.~\eqref{eq:inverseCDF}. The stochastic component of the model also ensures that, for each simulation, the generated extreme wind speeds appear at different time instants in the synthetic time series as a result of the different random initial phases \(\varphi_{k}^{\text{rnd}}\).      
\subsubsection{Autocorrelation Function}
\label{sec:orga4e4f22}
\begin{figure}[htbp]
\centering
\includegraphics[width=0.45\textwidth]{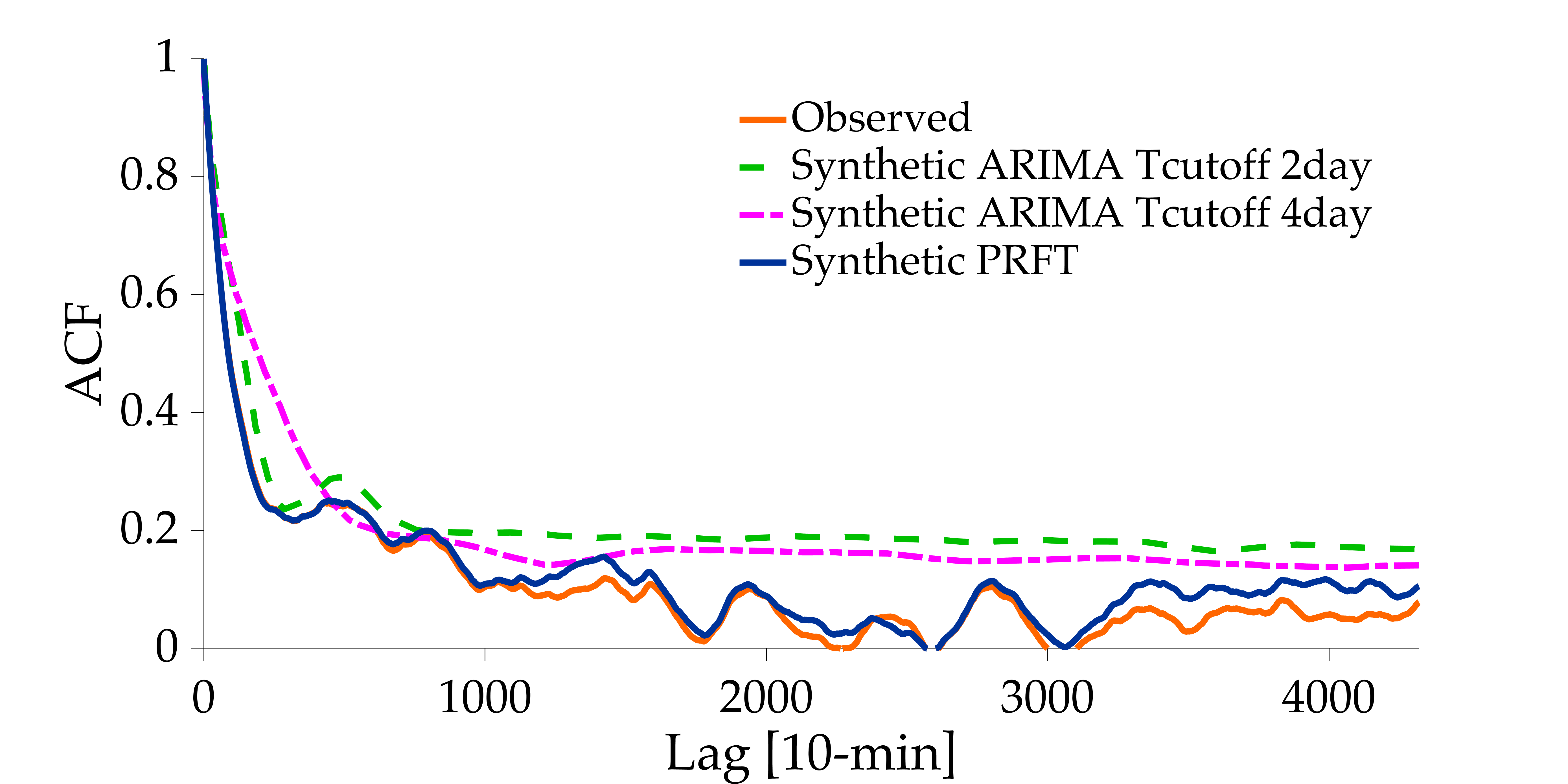}
\caption{\label{fig:orge31bc51}ACF agreement of the synthetic wind speeds simulated by the PRFT model and by the ARIMA-based model with the observed wind speeds at Näsudden.}
\end{figure}

A second level of comparison is conducted to assess the performance of the two models, PRFT and ARIMA-based, in simulating the temporal autocorrelation of the observed wind data at Näsudden. To do that, the autocorrelation function of the observed wind data is shown in Fig.~\ref{fig:orge31bc51} along with the synthetic wind data generated by the PRFT model, and the digitised ACF values of the ARIMA-simulated data for a time lag up to one month (4320 lags). For the latter, the figure shows the ACF values resulting from two different cut-off frequencies (\(1/T_{\text{cutoff}}\)), that determine different decompositions into low-frequency and high-frequency components in the ARIMA-based model.

A visual inspection reveals that the PRFT-simulated ACF follows very closely the autocorrelation of the wind speed measured at Näsudden throughout the whole range of analysed time lags. On the other hand, the ARIMA-simulated ACF that seems to yield a satisfactory match with the target autocorrelation profile (\(T_{\text{cutoff}} = 2\) days) can only reproduce the pattern of the observed ACF for approximately 500 lags, as its agreement degrades considerably at larger time lags. In addition, the ACF simulated by the ARIMA model shows a consistent bias throughout the calculated time lags. Yunus et al.~deem the ACF agreement given by their ARIMA-based model satisfactory, and comment that the agreement after 500 lags is less significant as the ACF of the observed wind speed is lower than 0.2. However, this comparison shows that the proposed PRFT model performs significantly better in reproducing the observed ACF at Näsudden, even for values lower than 0.2.      
\subsubsection{Power Spectral Density}
\label{sec:org57eba7b}
\begin{figure}[htbp]
\centering
\includegraphics[width=0.45\textwidth]{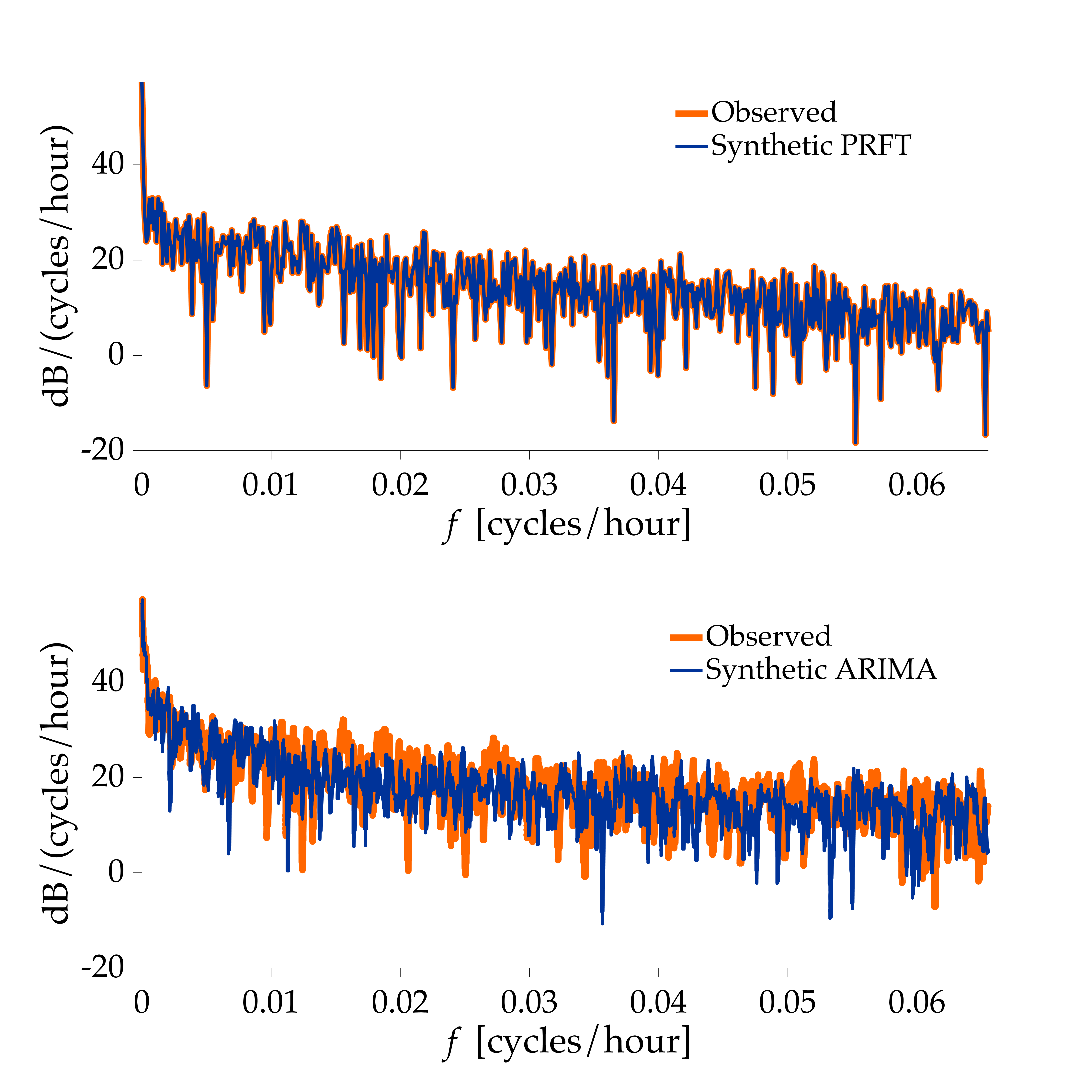}
\caption{\label{fig:org7dc1431}Periodogram agreement of synthetic wind speeds generated from Näsudden dataset.}
\end{figure}

Further information on how accurately the temporal autocorrelation and the periodic characteristics of the observed data are reproduced in the synthetic data can be inferred by performing a Fourier or spectral analysis on the observed and synthetised time series. For this reason, Yunus et al.~present the periodograms of the measured and the simulated wind-speed data. To provide a meaningful comparison with their investigation, the periodograms of the observed data and the synthetic data obtained from the PRFT model are calculated and shown in the top plot of Fig.~\ref{fig:org7dc1431} in the same units as in Yunus et al., namely dB/(cycles/hour). The bottom plot of Fig.~\ref{fig:org7dc1431} shows the digitised ARIMA-simulated periodogram along with the periodogram of the observed data calculated by Yunus et al. 

A visual comparison of the top plot of Fig.~\ref{fig:org7dc1431} reveals that the PRFT model can reproduce with very high accuracy the power spectral density of the observed wind-speed time series throughout the whole range of computed frequencies. The deviation of the PRFT-simulated periodogram with respect to the observed periodogram is given as the root-mean-square-error of the statistics, \emph{RMSE}\(_{\text{PER}}\), which yields a value of 0.0077.

In contrast, the bottom plot of Fig.~\ref{fig:org7dc1431} shows that the ARIMA-based model manages to capture only the decaying trend of the observed spectral content without matching accurately the magnitudes of the target periodogram. Yunus et al.~do not provide a metric that quantifies the observed deviation, therefore only a visual comparison is possible. Nevertheless, the results shown in this second level of comparison suffice to state that the proposed PRFT model outperforms the ARIMA-based model in reproducing both the temporal autocorrelation and the spectral content of the observed wind data at Näsudden.      
\subsubsection{User Interaction}
\label{sec:org41f6fe3}
In the modified ARIMA-based modelling procedure proposed by Yunus et al., user interaction is required throughout the process. During the first stage, the HF and LF components are obtained by performing a standard ARIMA modelling procedure. This entails user intervention to determine the proper combination of required transformations (i.e.~differencing and power transformation) to be applied to the observed wind data in order to identify the correct ARIMA model structure for the two components. In addition, a choice has to be made by the user for a suitable criterion to use for the determination of the transformation factor \(\nu\) that yields the best simulation results for each frequency component. Although the subsequent steps of the model identification can be automated, the modified ARIMA-based model also introduces shifting and limitation of the observed wind data before modelling to improve simulation results. The necessity to implement both steps is left to the user to judge based on the characteristics of the input time series. In a positive case, user interaction is required to determine suitable upper and lower limits and/or a constant offset value to apply to the input data that can be estimated by performing sensitivity analysis on the simulation results. Overall, the fine-tuning of the modified ARIMA-based model entails a high level of user interaction, that requires an expert time-series analyst with previous experience in ARIMA model identification to be performed effectively.    

In contrast, the PRFT model does not require user interaction throughout its operation and thus can be fully automated. In particular, no tuning of the model is needed to generate synthetic wind speeds from target data of different length and temporal resolution.  
\section{Conclusions}
\label{sec:org5376920}
In this paper, the PRFT modelling technique for the generation of realistic surrogate time series has been applied to wind speeds. The main contribution of this work is to show that this class of data-driven models can be suitably applied to generate surrogate data that conform to both the generally non-Gaussian probability density function and the power spectral density of a given wind-speed dataset. In addition, its performance in reproducing the diurnal and seasonal variations of the wind-speed has been analysed. It has been shown that, by its design, the proposed model yields a synthetic time series that matches exactly the PSD of the target dataset and converges to the target PDF with user-defined precision. Moreover, the stochastic component of the PRFT model (i.e.~its initial random phases) allows to generate multiple wind speed scenarios that share exactly the same PDF and PSD but represent different time evolutions of the synthetic wind speeds. This central feature can be used to generate multiple, realistic wind speed time series fully controlled by the user that can serve as input or training data for several applications.    

A comparison with two state-of-the-art models for the generation of surrogate wind data has been carried out to test the performance of the proposed PRFT model. The models selected for the comparison were the NHMC model of Xie et al.~and the ARIMA-based model of Yunus et al. The PRFT model has been applied to the same datasets used in the respective test cases of the selected models, and the comparison has been conducted in terms of the PDF, the ACF, and the PSD of the generated time series.

For both test cases, the PRFT-simulated wind speeds show a perfect reconstruction of both the PDF and the PSD of the target wind data. In terms of probability distribution, the PRFT model produces a marginally superior agreement with the target PDF compared to the NHMC model, whereas it yields a significantly better performance with respect to the ARIMA-based model. As for the power spectral density, the proposed model outperforms both the NHMC approach and the ARIMA-based model in reproducing the target PSD in the respective test cases. In particular, the PRFT-simulated PSD reproduces with high fidelity all the harmonics of the diurnal cycle in the Xie et al.~test case, while the NHMC-simulated PSD fails in getting the correct spectral amplitudes for some of those harmonics (3rd and 5th). The ACF analysis confirms this enhanced performance and shows that the PRFT-simulated wind speeds reproduce the same diurnal correlation of the observed wind data. A substantially superior performance is also shown by the PRFT model in the ACF analysis of the Yunus et al.~test case compared to the ARIMA-based model. Additionally, the proposed model sufficiently reproduces the seasonal variations of the wind data in the Xie et al.~test case showing the ability to capture such a nonstationary feature of the wind variation.

In addition, a user-interaction analysis has revealed that both the NHMC model and the ARIMA-based approach require user intervention to fine-tune the modelling process according to the characteristics of the input wind data. In contrast to them, the proposed PRFT model do not require any tuning from the user and shows identical performance when applied to datasets with different temporal resolutions (1 hour and 10 min, respectively), and different record lengths (10 years and 1 year, respectively). 

\bibliographystyle{unsrt}
\footnotesize{\bibliography{ms}}
\end{document}